\documentclass{article}

\usepackage{arxiv}

\usepackage[utf8]{inputenc} 
\usepackage[T1]{fontenc}    
\usepackage{hyperref}       
\usepackage[hyphenbreaks]{breakurl}
\usepackage{booktabs}       
\usepackage{amsfonts}       
\usepackage{nicefrac}       
\usepackage{microtype}      
\usepackage{lipsum}

\title{PUPILS pipeline: A flexible Matlab toolbox for eyetracking and pupillometry data processing}

\author{
  Helia Rela\~{n}o-Iborra \\
  Cognitive Systems\\
  Department of Applied Mathematics and Computer Science\\
  Technical University of Denmark\\
  2800 Kgs. Lyngby, Denmark \\
  \texttt{heliaib@dtu.dk} \\

   \And

    Per Bækgaard \\
Cognitive Systems\\
  Department of Applied Mathematics and Computer Science\\
  Technical University of Denmark\\
  2800 Kgs. Lyngby, Denmark \\
  \texttt{pgba@dtu.dk} \\
 
}

\begin{document}
\maketitle

\begin{abstract}

With the development of widely available commercial eye trackers, the use of eye tracking and pupillometry have become prevalent tools in cognitive research, both in industry and academia. However, dealing with
pupil recordings often proves challenging, as the raw data produced by the eye tracker is subject to noise and artifacts. With the PUPILS toolbox the authors wish to simplify the processing of pupil data for researchers. The toolbox receives the raw data extracted from the eye tracker output files and provides an
analysis of the relevant pupillary events (e.g., blinks and saccades), interpolates missing data and denoises the recorded traces. The final output of the toolbox is a cleaned up recording that can be easily exported for use in further analysis. 

\end{abstract}


\section{Introduction}\label{statement-of-need}

Changes in the pupil size have been linked to cognitive processes for decades \cite{Kahneman1966, Beatty1982a} for a review on early developments see \cite{Goldwater1972}. While originally recording the pupil's size and movements was a slow, manual and delicate process (e.g., see method sections in \cite{Hess1964, Kahneman1967}, in the last twenty years eye tracking and pupillometry have experienced a surge in popularity thanks to the rapid growth of commercially available eye trackers. However, state of the art trackers still produce data that is noisy and susceptible to have missing regions \cite{Hershman2018, Mathot2018}. The PUPILS pipeline is offered as a tool for scientist that are interested in analyzing pupil recordings, but might not want to spend a lot of time on cleaning the raw data output by their tracker. 

Thus, the user friendly PUPILS pipeline, built in Matlab \cite{MATLAB2019}, which receives the raw data as input and simply outputs a preprocessed recording ready to be further analyzed, allows researchers to obtain clean, anotated and usable data without investing too many resources. In addition to cleaning up the data, the toolbox classifies and stores different pupil events such as fixations, saccades and blinks, which provides a comprehensive description of the recordings that can be easily used for trial selection (e.g., to remove trials with a large proportion of blinks or isolate those portions of the recordings corresponding to fixations).

The PUPILS pipeline requires only two user-defined parameters in order for it to be able to process the recorded time series: the sampling frequency of the tracker, and the units used for recording the gaze coordinates (pixels, cm or mm). Nonwithstanding its simplicity, the toolbox also allows personalization of the algorithms and processing stages (e.g., users may choose their own parameters for minimum blink duration, saccade velocity threshold or interpolation regions, among other parameters) such that each individual user can best adapt the pipeline to the needs of their data. Furthermore, its modular structure allows for the relatively simple inclusion of user-defined functions and processing stages, thus making it a flexible framework that can work with different experimental and analysis designs.

The pipeline has been validated for data recorded with low (e.g., 60 Hz) and high sample frequencies (e.g., 500 Hz), thus should be compatible with most of the state of the art trackers, across different price and quality ranges.

\hypertarget{pipeline-structure}{%
\section{Pipeline structure}\label{pipeline-structure}}

The main function of the pipeline is \texttt{processPupilData.m} which receives the data to be processed and a structure of options, that must include the compulsory fields of sampling frequency and measurement units, and any other optional fields the user might want to specify. Option fields not specified by the user will run in default mode. That is to say, a user might simply call this function and will obtain the cleaned pupil response as well as event and data quality information. The output consist on a data frame that contains the original data with appended annotations about blink information, saccade information, interpolated data and denoised data and an information structure which contains metadata regarding eye movement events and quality of the pre-processed data.

This function provides the skeleton of the PUPILS pipeline flexible modular design and it is build as consecutive calls to the different processing stages of the processing pipeline: blink detection, saccade detection, data interpolation and smoothing. In this way, if a user would want to include a different processing module (e.g., a different type of denoising approach or a fixation detection algorithm), it could be easily incorporated to the \texttt{processPupilData.m} structure.

The blink detection algorithm included in the PUPILS pipeline defines a blink region as those regions of the recording where the measured pupil size is smaller than three standard deviations lower than the mean of the entire recording, as recommended by \cite{Winn2018}.

In its default mode, the saccade detection is performed using the algorithm proposed by \cite{Duchowski2002}, which defines the angular velocity as:

\begin{equation}
v = \frac{\theta}{\Delta t}
\end{equation}

where time increment \(\Delta t\) is defined in the PUPILS pipeline as the duration of five time samples (i.e., \(5 \cdot \frac{1}{f_{s}})\)), with \(f_{s}\) being the sampling frequency. \(\theta\) represents the visual angle calculated as:

\begin{equation}
\theta = \cos^{-1} \frac{(\mathbf{P}-\mathbf{C})\cdot (\mathbf{Q}-\mathbf{C})}{\parallel (\mathbf{P}-\mathbf{C}) \parallel \parallel (\mathbf{Q}-\mathbf{C}) \parallel}
\end{equation}

In the PUPILS pipeline, \(\mathbf{C}\) is considered the center of coordinates (i.e., the head position) and thus assigned the value {[}0, 0, 0{]}. \(\mathbf{P}\) and \(\mathbf{Q}\), are positional vectors built using the recorded \(x\) and \(y\) gaze coordinates output by the tracker for each considered sample and its 5-sample predecessor. Both vectors share their \(z\)-coordinate, a constant corresponding to the user-defined distance from the head to the screen or visual target. Note that the distance to the screen needs to be expressed in the same units as the \(x\) and \(y\) coordinates recorded by the tracker.

In the default mode of the PUPILs pipeline a saccade is detected when the calculated velocity is over a velocity threshold of \(30^{\mathrm{\circ / s} }\), as suggested by \cite{Holmqvist2011}. The velocity threshold may be defined by the user, and this is indeed highly recommended (e.g., lower thresholds may be required for experiments where detecting slow saccadic movements is key). The velocity values output by the PUPILS pipeline, might give a good indication of which velocity threshold is ideal for the data at hand.

Missing data and blink portions are linearly interpolated. The region of interpolation includes and additional 50 ms prior and 150 ms after the detected missing portions to avoid measurement artifacts related to blinks \cite{Winn2018}. However, both pre- and post-interval regions can be defined by the user. The final denoised signal is obtained by low pass filtering the interpolated traces (if not user defined, cut-off frequency of the filter is 10 Hz).

\hypertarget{data-format}{%
\section{Data format}\label{data-format}}

The input dataframe has rows corresponding to temporal samples and columns with different data categories. The minimum dataframe should include: time stamp, x-coordinate, y-coordinate and pupil size, \emph{in that order}. In addition, x and y velocity values may be included. Note that, currently, the pipeline works for one eye at a time. So when data for both eyes needs to be analysed two input structures must be created and the pipeline must be called twice.

The pipeline outputs a dataframe that contains the original data with appended columns that contain, in the following order, blink information, saccade information, the interpolated data and the denoised (i.e., smoothed) data. If velocity data was not originally provided, the calculated velocity will be appended directly after the original dataframe, as the fifth column, right before the blink information.

\section{Availability}\label{rep-info}

Code for the pipeline is offered to the research community at: \sloppy
\url{https://gitlab.gbar.dtu.dk/heliaib/PUPILS-preprocessing-pipeline}. The repository can be cloned using https or SSH protocols. Alternatively, the contents of the repository can also be downloaded using the GitLab interface.

\section*{Acknowledgements}\label{acknowledgements}

The authors would like to acknowledge Tanya Bafna for comments and
suggestions on the pipeline.

\section*{References}\label{references}


\end{document}